\documentclass[a4paper,10pt,twoside]{article}
\usepackage{latexsym}
\usepackage{amsmath}
\usepackage{amssymb}
\usepackage{amscd}

\newlength{\minitwocolumn}
\makeatletter
\@addtoreset{equation}{section}
\makeatother

\newtheorem{thm}{Theorem}
\newtheorem{prop}[thm]{Proposition}

\title{Quantum Knizhnik-Zamolodchikov equations of level $0$ \\ 
and form factors in SOS model%
}

\author{Yas-Hiro Quano\thanks{
E-mail: quanoy@suzuka-u.ac.jp}
}

\date{\it 
Department of Clinical Engineering, 
Suzuka University of Medical Science \\
Kishioka-cho 1001-1, Suzuka 510-0293, Japan}

\begin{document}

\maketitle

\begin{abstract}
The cyclic SOS model is considered on the basis of Smirnov's 
form factor bootstrap approach. Integral solutions to the 
quantum Knizhnik-Zamolodchikov equations of level $0$ 
are presented. 
\end{abstract}

\section{Intorduction}

In a previous paper \cite{qXYZ} we presented integral 
formulae for correlation functions of both cyclic SOS 
and antiferromagnetic XYZ models, 
by directly solving the bootstrap equations. 
The bootstrap equations, the master equations of 
correlation functions in integrable models, can be 
derived on the basis of the CTM bootstrap \cite{JMN}. 
In the present paper, we are interested in form factors 
of the elliptic integrable model. Form factors are very 
important objects, because all physical quantities can be 
expressed in terms of form factors, in principle. 

Through the study of form factors of the 
sine-Gordon model, Smirnov \cite{Smbk} 
found three axioms as sufficient conditions for the 
local commutativity of local fields in the model. 
These three axioms consist of (i) $S$-matrix symmetry, 
(ii) the cyclicity condition, and (iii) the annihilation pole 
condition. Furthermore, Smirnov observed in Ref. \cite{Sm1} 
that the first two axioms imply the quantum 
Knizhnik-Zamolodchikov equation \cite{FR} of level $0$. 

Form factors were originally defined as 
matrix elements of local operators. However, 
we refer to the objects that satisfy 
Smirnov's three axioms as `form factors'. 
In this sense, we wish to construct integral formulae 
for cyclic SOS form factors in this paper. 
This is a preliminary study, in which 
we make preparations to construct those of the XYZ 
form factors. 

There are several ways to address the problem of constructing 
form factors of integrable models. 
In the vertex operator approach, \cite{JMbk,Lu} 
form factors can be constructed 
as traces of products of type II vertex operators 
and some local operators. Lashkevich \cite{La} used the 
vertex-face correspondence \cite{3-bu/2} to construct 
a free field representation of the type II vertex operators 
of the XYZ model in terms of those of 
the eight-vertex SOS model \cite{LuP,MW}, using an 
insertion of a non-local 
tail operator. The type I vertex operators of 
the XYZ model were constructed in Ref. \cite{LaP}. These 
operators are relevant for the expression of the local operators. 
Shiraishi \cite{Shi} directly constructed 
the type II and type I vertex operators of the eight-vertex 
model as intertwiners of the $q$-deformed Virasoro algebra 
\cite{SKAO}, without using the vertex-face correspondence. 

After Smirnov's pioneering works on the 
sine-Gordon model \cite{Smbk,S1,BBS2}, an axiomatic approach was 
applied to the XXZ form factors \cite{JKMQ,KMQ,affine}, and 
the $SU(2)$-invariant Thirring model \cite{NT,N3}. 
In this paper we wish 
to address the problem using the same approach. 

In Refs. \cite{Matsuo,TV}, Jackson-type integral 
formulae were constructed. The relations among those approaches 
(the vertex operator formalism, the axiomatic approach, 
and the Jackson integral representation approach) were discussed 
in Ref. \cite{NPT}. 
A rigorous method to obtain correlation functions 
and form factors was developed in Refs. \cite{KMT0,KMST1} on 
the basis of the algebraic Bethe ansatz method. Unfortunately, 
for technical reasons, this rigorous method has not yet 
been applied to elliptic integrable models, such as 
the SOS and XYZ models. 

Integral formulae of the XYZ model 
form factors can be obtained from those of the SOS model 
form factors by using the vertex-face correspondence. 
As a preliminary task to be performed 
for this purpose, we present integral 
solutions to the quantum Knizhnik-Zamolodchikov equation of 
level $0$ associated with the cyclic SOS model. 

The rest of the present paper is organized as follows. 
In \S 2 we derive quantum Knizhnik-Zamolodchikov equations 
of level $0$ that form factors 
in the cyclic SOS model should satisfy. In \S 3 we 
construct integral formulae for SOS form factors as solutions to 
level $0$ quantum Knizhnik-Zamolodchikov equations. 
In \S 4 we give some concluding remarks. In Appendix 
A we give a proof of Proposition \ref{prop:qKZ}. 

\section{Smironv's axioms in the SOS model}

\subsection{Boltzmann weights of the cyclic SOS model}

The eight-vertex 
SOS model is a face model \cite{3-bu/2} which is 
defined on a square lattice with a site variable 
$k_j \in \mathbb{Z}$ attached to each site $j$. 
We call $k_j$ the local state or height 
and impose the condition 
that heights of adjoining sites differ by $1$. 
Local Boltzmann weights are assumed 
to be functions of the spectral parameter $\zeta =x^{-u}$ and 
are denoted by 
$$
\displaystyle W\left[ \left. \begin{array}{cc} 
c & d \\ b & a \end{array} \right| \zeta \right]=
W\left[ \left. \begin{array}{cc} 
c & d \\ b & a \end{array} \right| u \right], 
$$
which is defined for a state configuration 
$(a,b,c,d)$ ordered clockwise 
from the SE corner around the face. 
For fixed $x$ and $r$ satisfying $0<x<1$ and $r>1$, 
the nonzero Boltzmann weights are given as follows: 
\begin{eqnarray}
\displaystyle W\left[ \left. \begin{array}{cc} 
k\pm 2 & k\pm 1 \\ k\pm 1 & k \end{array} \right| u \right]
& = & \dfrac{1}{\kappa (u)}, \nonumber \\
\displaystyle W\left[ \left. \begin{array}{cc} 
k & k\pm 1 \\ k\pm 1 & k \end{array} \right| u \right]
& = & \displaystyle \dfrac{1}{\kappa (u)}
\frac{[1]\{k\pm u\}}{[1-u]\{k\}}, \nonumber \\
\displaystyle W\left[ \left. \begin{array}{cc} 
k & k\mp 1 \\ k\pm 1 & k \end{array} \right| u \right]
& = & \displaystyle \dfrac{1}{\kappa (u)}
\frac{[u]\{k\pm 1\}}{[1-u]\{k\}}. 
\label{eq:HY}
\end{eqnarray}
Here, $0<u<1$ (Regime III), and we use the following symbols: 
$$
{[}u{]}
=\displaystyle x^{\frac{u^2}{r}-u}\Theta_{x^{2r}}(x^{2u}), 
~~~~ 
\{ u\} =\displaystyle x^{\frac{u^2}{r}-u}
\Theta_{x^{2r}}(-x^{2u}), 
$$
where
$$
\Theta_{p}(z):=(z; p)_\infty 
(pz^{-1}; p)_\infty (p, p)_\infty , ~~~~
(a;p_1,\cdots,p_n)_\infty=
\displaystyle \prod_{k_i\geqslant 0}(1-ap_1^{k_1}\cdots p_n^{k_n}). 
$$
The Jacobi theta functions are defined by 
$$
\theta_1 (u;\tau )=\sqrt{-1}q^{\frac{1}{4}}
e^{-\sqrt{-1}\pi u} 
\Theta_{q^2} (e^{2\sqrt{-1}\pi u}), \;\;\;\;
\theta_4 (u;\tau )=
\Theta_{q^2} (q e^{2\sqrt{-1}\pi u}), 
$$
where $q=\exp (\sqrt{-1}\pi\tau )$. Note that 
$$
\theta_{1} (\tfrac{u}{r}; 
\tfrac{\pi \sqrt{-1}}{\epsilon r})
=\sqrt{\tfrac{\epsilon r}{\pi}}\exp \left( 
-\tfrac{\epsilon r}{4} \right) [u], 
~~~~\theta_{4} (\tfrac{u}{r}; 
\tfrac{\pi \sqrt{-1}}{\epsilon r})
=\sqrt{\tfrac{\epsilon r}{\pi}}\exp \left( 
-\tfrac{\epsilon r}{4} \right) \{u\}. 
$$
The normalization factor 
$$
\kappa (\zeta)= \displaystyle\zeta^{\frac{r-1}{r}}
\frac{(x^4 z ; x^4 , x^{2r})_{\infty}
      (x^{2} z^{-1} ; x^4 , x^{2r})_{\infty}
      (x^{2r} z ; x^4 , x^{2r})_{\infty}
      (x^{2r+2} z^{-1} ; x^4 , x^{2r})_{\infty}}
     {(x^{4} z^{-1} ; x^4 , x^{2r})_{\infty}
      (x^2 z ; x^4 , x^{2r})_{\infty}
      (x^{2r} z^{-1} ; x^4 , x^{2r})_{\infty}
      (x^{2r+2}z ; x^4 , x^{2r})_{\infty}}, 
$$
where $z=\zeta^2$, 
is chosen such that the partition function per site 
is unity. 

The Boltzmann weights (\ref{eq:HY}) satisfy the face-type 
Yang-Baxter equation: 
\begin{eqnarray}
&& \displaystyle \sum_{g} 
W\left[ \left. 
\begin{array}{cc} e & f \\ g & a \end{array} \right| 
u_1 -u_2 \right] W\left[ \left. 
\begin{array}{cc} d & e \\ c & g \end{array} \right| 
u_1 \right]
W\left[ \left. 
\begin{array}{cc} c & g \\ b & a \end{array} \right| 
u_2 \right] \nonumber \\
&= & \displaystyle \sum_{g} 
W\left[ \left. 
\begin{array}{cc} d & e \\ g & f \end{array} \right| 
u_2 \right]
W\left[ \left. 
\begin{array}{cc} g & f \\ b & a \end{array} \right| 
u_1 \right]
W\left[ \left. 
\begin{array}{cc} d & g \\ c & b \end{array} \right| 
u_1 -u_2 \right]. 
\label{eq:WYBE}
\end{eqnarray}
We can construct $R(u,\lambda )$ from $W$ such that 
$R(u,\lambda )$ satisfies the dynamical Yang-Baxter equation 
\cite{dYBE}.

\subsection{Vertex operators and Smirnov's axioms} 

Let $p=(k_1 , k_2 , k_3 , \cdots )$, with 
$|k_{j+1}-k_j|=1$ ($j=1,2,3,\cdots$), be an admissible 
path, and let 
${\cal H}_{l,k}^{(i)}$ ($i=0,1$) be the space of admissible 
paths satisfying the initial condition $k_1 =k$ and 
the following boundary conditions: 
$$
k_j =\left\{ \begin{array}{lll} 
l & \mbox {if $j\equiv 1-i$} & \mbox{(mod $2$)}, \\
l+1 & \mbox {if $j\equiv i$} & \mbox{(mod $2$)}. 
\end{array} \right. ~~~~ (j\gg 1) 
$$
The type II vertex operator in the SOS model, 
$\Psi^{*(1-i,i)}(\zeta )_l^{l'}$ 
($l'=l\pm 1$), acts as \cite{LuP,MW} 
\begin{equation}
\Psi^{*(1-i,i)}(\zeta )_l^{l'} : 
{\cal H}_{l,k}^{(i)}\rightarrow {\cal H}_{l',k}^{(1-i)} 
\end{equation}
and satisfies the commutation relation 
\begin{equation}
\Psi^{*}(\zeta_1 )_b^{c}\Psi^{*}(\zeta_2 )_a^{b}
=\sum_{d}\Psi^{*}(\zeta_2 )_d^{c}
\Psi^{*}(\zeta_1 )_a^{d}
W'\left[ \left. 
\begin{array}{cc} c & d \\ b & a \end{array} \right| 
\zeta_1 /\zeta_2 \right], 
\end{equation}
where 
$$
W'\left[ \left. 
\begin{array}{cc} c & d \\ b & a \end{array} \right| 
\zeta \right]=-W\left[ \left. 
\begin{array}{cc} c & d \\ b & a \end{array} \right| 
\zeta \right] \left. \makebox{\rule[-4mm]{0pt}{11mm}} 
\right|_{r\mapsto r-1}.
$$ 

In this subsection, we consider form factors 
in the cyclic SOS model of the following form: 
\begin{equation}
F_n^{(i)}({\cal O}; \zeta_1 , \cdots , \zeta_{2n})_{
ll_1 \cdots l_{2n-1}l}:=
\langle \hat{{\cal O}} 
\Psi^{*(i,1-i)}(\zeta_{2n})_{l_{2n-1}}^l 
\cdots \Psi^{*(1-i,i)}(\zeta_1 )_l^{l_1} 
\rangle_i, 
\end{equation}
where $|l_j -l_{j-1}|=1$ ($1\leqslant j\leqslant 2n$), 
with $l_0 =l_{2n}=l$. 
Let 
$$
F^{(\sigma )}_n ({\cal O}; \zeta )_{
ll_1 \cdots l_{2n-1}l}
=F_n^{(0)}({\cal O};\zeta)_{ll_1 \cdots l_{2n-1}l}+\sigma 
F_n^{(1)}({\cal O}; \zeta)_{ll_1 \cdots l_{2n-1}l}. ~~~~ 
(\sigma =\pm 1)
$$
Then the following three axioms for the cyclic SOS model hold. 

\noindent{\it 1. $W'$--symmetry}
\begin{eqnarray}
&&F^{(\sigma )}_{n} 
({\cal O}; \cdots,\zeta_{j+1},\zeta_j,\cdots)_{\cdots 
l_{j-1}l_jl_{j+1} \cdots} \label{eq:W'-symm-comp} \\[4mm]
&=&\displaystyle\sum_{l'_j} 
W'\left[ \left. \begin{array}{cc} 
l_{j+1} & l'_j \\ l_j & l_{j-1} \end{array} \right| 
\zeta_j/\zeta_{j+1} \right]
F^{(\sigma )}_{n} 
({\cal O}; \cdots,\zeta_j,\zeta_{j+1},\cdots)_{\cdots 
l_{j-1}l'_jl_{j+1} \cdots}. \nonumber
\end{eqnarray}
\noindent{\it 2. Cyclicity}
\begin{equation}
F^{(\sigma )}_{n} 
({\cal O}; \zeta', x^{-2}\zeta_{2n})_{
ll_1\cdots l_{2n-1}l}
=\sigma F^{(\sigma )}_{n} 
({\cal O}; \zeta_{2n}, \zeta')_{
l_{2n-1}ll_1\cdots l_{2n-1}}, 
\label{eq:F'cyc-comp}
\end{equation}
where $\zeta' =(\zeta_1 , \cdots , \zeta_{2n-1})$. 

\noindent{\it 3. Annihilation pole condition}
\begin{eqnarray}
&& \underset{\zeta_{2n}=\varepsilon x^{-1}\zeta_{2n-1}}{\rm Res}\;
F^{(\sigma )}_n ({\cal O}; \zeta )_{ll_1\cdots l'l}
\dfrac{d\zeta_{2n}}{\zeta_{2n}} =
\{ l' \}' \left( \makebox{\rule[-4mm]{0pt}{11mm}}
\delta_{l_{2n-2}l} 
F^{(\varepsilon \sigma )}_{n-1} 
({\cal O}; \zeta'')_{ll_1\cdots l_{2n-3}l} \right. 
\label{eq:res'-cond} \\[3mm]
&&\left. -\sigma\varepsilon \displaystyle\sum_{
l'_2\cdots l'_{2n-2}} F^{(\varepsilon \sigma )}_{n-1} 
(\zeta' )_{l'l'_2\cdots l'_{2n-2}l'}
\prod_{j=1}^{2n-2} 
W'\left[ \left. \begin{array}{cc} 
l'_{j+1} & l'_j \\ l_j & l_{j-1} \end{array} \right| 
\zeta_{2n-1}/\zeta_{j} \right] \right), \nonumber 
\end{eqnarray}
where $l_0 =l$, $l'_{2n-1}=l'$ on the RHS of (\ref{eq:res'-cond}). 

The first two axioms, (\ref{eq:W'-symm-comp}) and (\ref{eq:F'cyc-comp}), 
imply the quantum Knizhnik-Zamolodchikiv equations of 
level $0$: 
\begin{eqnarray}
&&F^{(\sigma )}_n ({\cal O}; \zeta_1 , \cdots , 
x^2 \zeta_j , \cdots , \zeta_{2n} )_{l_0 l_1\cdots l_{2n-1}l_0} 
\nonumber \\[3mm]
&=& \sigma\displaystyle\sum_{l'_1 \cdots l'_{j-1} l'_{j+1} \cdots l'_{2n}} 
W'\left[ \left. \begin{array}{cc} 
l_{j} & l'_{j-1} \\ l_{j-1} & l_{j-2} \end{array} \right| 
\frac{x^2 \zeta_j}{\zeta_{j-1}} \right] 
\prod_{k=1}^{j-2} W'\left[ \left. \begin{array}{cc} 
l'_{k+1} & l'_{k} \\ l_{k} & l_{k-1} \end{array} \right| 
\frac{x^2 \zeta_j}{\zeta_{k}} \right] \nonumber \\[5mm]
&\times & \displaystyle W'\left[ \left. \begin{array}{cc} 
l'_{1} & l'_{2n} \\ l_{0} & l_{2n-1} \end{array} \right| 
\frac{\zeta_j}{\zeta_{2n}} \right] 
\prod_{k=j+1}^{2n} W'\left[ \left. \begin{array}{cc} 
l'_{k+1} & l'_{k} \\ l_{k} & l_{k-1} \end{array} \right| 
\frac{\zeta_j}{\zeta_{k}} \right] \nonumber \\[5mm]
&\times & F^{(\sigma )}_n ({\cal O}; \zeta_1 , \cdots , 
\zeta_j , \cdots , \zeta_{2n} )_{l'_1\cdots l'_{j-1}l_jl'_{j+1}\cdots 
l'_{2n} l'_1}. 
\label{eq:qKZ}
\end{eqnarray}
In what follows, we consider only the first two axioms. 

\section{Integral formulae for the cyclic SOS form factors}

Let us recall $z_j=\zeta_j^2 =x^{-2u_j}$. Set 
\begin{equation}
 F_{n}^{(\sigma)}({\cal O}; \zeta)_{ll_1\cdots l_{2n-1}l}
=c_n \prod_{1\leqslant j< k \leqslant 2n} 
\zeta_j^{-\frac{r}{r-1}} g^*(z_j/z_k)
\times \overline{F}_{n}^{(\sigma)}
({\cal O}; \zeta)_{ll_1\cdots l_{2n-1}l}. 
\label{eq:G-bar}
\end{equation}
Here, $c_n$ is a constant, 
and the function $g^*(z)$ satisfies
\begin{equation}
\kappa^* (\zeta )=-\kappa (\zeta ; x, x^{2(r-1)})
=\zeta^{-\frac{r}{r-1}} 
\frac{g^*(z)}{g^*(z^{-1})}, ~~~~ 
g^* (z)=g^* (x^4 z^{-1}). 
\label{eq:g*-prop}
\end{equation}
The explicit form of $g^*(z)$ is as follows: 
\begin{equation}
g^* (z)=\frac{
\{ z\}'_\infty \{ x^4 z^{-1}\}'_\infty 
\{ x^{2r+2}z\}'_\infty \{ x^{2r+6}z^{-1}\}'_\infty}
{\{ x^2 z\}'_\infty \{ x^6 z^{-1}\}'_\infty 
\{ x^{2r}z\}'_\infty \{ x^{2r+4}z^{-1}\}'_\infty}, \;\,
\{ z\}'_\infty =(z; x^4 , x^4 , x^{2(r-1)}). 
\label{eq:g*}
\end{equation}

In order to present our integral formulae for 
$\overline{F}_{n}^{(\sigma)}({\cal O}; \zeta )$, let us 
prepare some notation. Let 
\begin{equation}
A_\pm :=\{ a|l_a =l_{a-1}\pm 1, 
\,\,1\leqslant a\leqslant 2n \}. 
\label{eq:df-A}
\end{equation}
Then, the number of elements of $A_\pm$ is equal to $n$, 
because we now set $l_{2n}=l_0 =l$. 
We often use the abbreviated notation 
$(w)=(w_{a_1}, \cdots , w_{a_n})$ and 
$(w')=(w_{a_1}, \cdots , w_{a_{n-1}})$ 
for $a_j \in A_\pm$ 
such that $a_1 <\cdots <a_n$. Let us define the 
meromorphic function 
\begin{equation}
Q_{n}(w|\zeta )_{ll_1\cdots l_{2n-1}l}
=\displaystyle\prod_{a\in A_-} 
\frac{\{v_a -u_a -\tfrac{3}{2}-l_a \}'}{
[u_a -v_a +\tfrac{3}{2}]'} \left( 
\prod_{j=1}^{a-1} \frac{[v_a -u_j -\tfrac{1}{2}]'}{
[u_j -v_a + \tfrac{3}{2}]'} \right) 
\prod_{a,b\in A_-\atop a<b} 
[v_a -v_b +1]', 
\label{eq:df-QF}
\end{equation}
where $w_a =x^{-2v_a}$, and 
$$
{[}u{]'}
=\displaystyle 
x^{\frac{u^2}{r-1}-u}\Theta_{x^{2(r-1)}}(x^{2u}), 
~~~~ 
\{ u\}' =\displaystyle x^{\frac{u^2}{r-1}-u}
\Theta_{x^{2(r-1)}}(-x^{2u}). 
$$
Here, we should note that 
the structure of the expression (\ref{eq:df-QF}) is 
quite similar to that given in Refs. \cite{NT,NPT}. 

We wish to find integral formulae of the form 
\begin{equation}
\overline{F}_{n}^{(\sigma)} 
({\cal O}; \zeta)_{ll_1\cdots l_{2n-1}l}
=\prod_{a\in A_-}\oint_{C_a} 
\dfrac{dw_a}{2\pi\sqrt{-1}w_a} 
\Psi_{n}^{(\sigma)} (w|\zeta )
Q_{n}(w|\zeta)_{ll_1\cdots l_{2n-1}l}P_{\cal O}(w). 
\label{eq:G-form}
\end{equation}
Here, $P_{\cal O}(w)$ is an antisymmetric holomorphic 
function of $w_{a_1}, \cdots , w_{a_n}$. 
The kernel has the form 
\begin{equation}
\Psi^{(\sigma)}_n (w| \zeta )=
\vartheta^{(\sigma)}_n (w | \zeta )
\prod_{a\in A_-}\prod_{j=1}^{2n} 
x^{-\frac{(v_a -u_j)^2}{2(r-1)}} 
\psi \Bigl(\frac{w_{a}}{z_j}\Bigr) 
\prod_{1\leqslant j<k\leqslant 2n} 
x^{-\frac{(u_j -u_k)^2}{4(r-1)}},
\label{eq:df-Psi}
\end{equation}
where
\begin{equation}
\psi(z)=
\frac{(x^{2r+1}z;x^4,x^{2(r-1)})_{\infty}
(x^{2r+1}z^{-1};x^4,x^{2(r-1)})_{\infty}}
{(xz;x^4,x^{2(r-1)})_{\infty}(xz^{-1};x^4,x^{2(r-1)})_{\infty}}. 
\label{eq:df-psi}
\end{equation}

For the function 
$\vartheta^{(\sigma)}_n (w|\zeta )$, 
we assume the following: 
\begin{itemize}
\item
it is anti-symmetric and holomorphic
in the variables $w_a \in \mathbb{C} \backslash \{0\}$,
 
\item
it is symmetric and meromorphic
in the variables $\zeta_j\in \mathbb{C} \backslash \{0\}$,
 
\item
it has the two transformation properties 
\begin{eqnarray}
\displaystyle\frac
{\vartheta^{(n)}_\sigma (w |\zeta', x^2 \zeta_{2n})}
{\vartheta^{(n)}_\sigma (w | \zeta )}&=& 
\sigma x^{-2n+\frac{2n-1}{r-1}} 
\displaystyle\prod_{a\in A}w_{a}^{-1}
\prod_{j=1}^{2n} \zeta_j, 
\label{eq:z-sym}\\
\displaystyle \frac
{\vartheta^{(n)}_\sigma (w', x^4 w_{a_n}| \zeta )}
{\vartheta^{(n)}_\sigma (w | \zeta )}
&=& x^{-4n} \displaystyle\prod_{j=1}^{2n} 
\frac{ w_{a_n}}{z_j }.
\label{eq:x-sym}
\end{eqnarray}
\end{itemize}
The function $\vartheta_{n}^{(\sigma)}(w|\zeta )$
is otherwise arbitrary, and the choice of 
$\vartheta_{n}^{(\sigma)}(w|\zeta )$ 
corresponds to that of solutions to 
(\ref{eq:W'-symm-comp})--(\ref{eq:res'-cond}). 
The transformation properties of 
$\vartheta_{n}^{(\sigma)}(w|\zeta )$ imply
\begin{eqnarray}
\frac{\Psi_{n}^{(\sigma)}(w|\zeta', x^{-2} \zeta_{2n})}
{\Psi_{n}^{(\sigma)}(w|\zeta )}
&=&\sigma\prod_{j=1}^{2n} 
\left( \frac{\zeta_j}{\zeta_{2n}}\right)^{\frac{r}{r-1}}
\prod_{a\in A_-}\frac{[v_a -u_{2n} -\frac{1}{2}]'}
{[u_{2n}-v_a +\frac{3}{2}]'}, 
\label{eq:trPsi1} \\
\frac{\Psi_{n}^{(\sigma)}(w',x^{-4}w_{a_n}|\zeta )}
{\Psi_{n}^{(\sigma)}(w|\zeta )}
&=&\prod_{j=1}^{2n}\frac{[u_{j}-v_{a_n}-\frac{1}{2}]'}
{[v_{a_n}-u_{j}+\frac{3}{2}]'}.
\label{eq:trPsi2}
\end{eqnarray}

The integrand may have poles at the values 
\begin{equation}
w_a =\left\{ 
\begin{array}{ll}
x^{\pm (1+4k_1+2(r-1)k_2 )}z_j, & (1\leqslant j\leqslant 2n, 
k_1 , k_2\in \mathbb{Z}_{\geqslant 0}) \\
x^{-3+2(r-1)k}z_j. & (1\leqslant j\leqslant 2n, 
k\in \mathbb{Z}) 
\end{array} \right. 
\label{eq:pole-position}
\end{equation}
We choose the integration contour $C_a$ with 
respect to $w_a$ ($a\in A_-$) to be along 
a simple closed curve oriented counter-clockwise that 
encircles the points $x^{1+4n_1+2(r-1)n_2}z_j$ 
$(1\leqslant j \leqslant 2n , 
n_1 , n_2\in \mathbb{Z}_{\geqslant 0})$ 
and $x^{-3+2(r-1)n_3}z_j$ $(1\leqslant j \leqslant 2n , 
n_3\in \mathbb{Z}_{\geqslant 0})$, 
but not $x^{-1-4n_1-2(r-1)n_2} z_j$ 
$(1\leqslant j \leqslant 2n, 
n_1 , n_2\in \mathbb{Z}_{\geqslant 0})$ 
nor $x^{-3-2(r-1)n_3}z_j$ $(1\leqslant j \leqslant 2n , 
n_3\in \mathbb{Z}_{>0})$. Thus, the contour $C_a$ actually 
depends on the variables $z_j$ in addition to $a$, and therefore 
strictly, it should be written $C_a (z)=C_a (z_1, \cdots , z_{2n})$. 
The LHS of (\ref{eq:F'cyc-comp}) represents the analytic 
continuation with respect to $\zeta_{2n}$. 

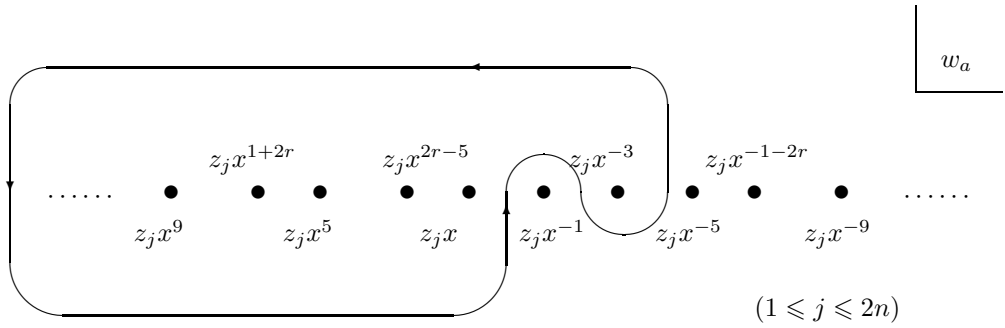
\unitlength 1.65mm
\begin{picture}(140,35)
\put(-37,-10){
\put(55,25){\oval(40,20)[b]}
\put(38,35){\line(1,0){34}}
\put(38,32){\oval(6,6)[tl]}
\put(35,32){\vector(0,-1){7}}
\put(67,25){\circle*{1}}
\put(72,25){\circle*{1}}
\put(60,25){\circle*{1}}
\put(55,25){\circle*{1}}
\put(48,25){\circle*{1}}
\put(38,24.1){$\cdots\cdots$}
\put(78,25){\circle*{1}}
\put(84,25){\circle*{1}}
\put(90,25){\circle*{1}}
\put(95,25){\circle*{1}}
\put(102,25){\circle*{1}}
\put(107,24.1){$\cdots\cdots$}
\put(68,21){$z_j x$}
\put(65,27){$z_j x^{2r-5}$}
\put(57,21){$z_j x^5$}
\put(51,27){$z_j x^{1+2r}$}
\put(45,21){$z_j x^{9}$}
\put(76,21){$z_j x^{-1}$}
\put(80,27){$z_j x^{-3}$}
\put(87,21){$z_j x^{-5}$}
\put(91,27){$z_j x^{-1-2r}$}
\put(99,21){$z_j x^{-9}$}
\put(75,19){\vector(0,1){5}}
\put(78,25){\oval(6,6)[t]}
\put(84.5,25){\oval(7,7)[b]}
\put(88,25){\line(0,1){7}}
\put(85,32){\oval(6,6)[tr]}
\put(85,35){\vector(-1,0){13}}
\put(95,15){\normalsize($1\leqslant j\leqslant 2n$)}
\put(110,35){$w_a$}
\put(108,33){\line(0,1){7}}
\put(108,33){\line(1,0){7}}
}
\end{picture}

We are now in a position to state the following main proposition 
of the present paper: 

\begin{prop} Assume the properties of 
the function $\vartheta^{(n)}_\sigma$ given 
below (\ref{eq:df-psi}) and the integration contour 
$C_a$ given below (\ref{eq:pole-position}). 
Then the integral formulae 
(\ref{eq:G-bar}) and (\ref{eq:G-form}) 
with (\ref{eq:g*}), (\ref{eq:df-QF}), 
(\ref{eq:df-Psi}) and (\ref{eq:df-psi}) 
solve the quantum Knizhnik-Zamolodchikov equation 
(\ref{eq:qKZ}). 
\label{prop:qKZ}
\end{prop}

We give a proof of Proposition \ref{prop:qKZ} 
in Appendix A. 

\section{Concluding remarks}

In this paper we have constructed integral formulae for 
form factors of the cyclic SOS model as solutions to 
the quantum Knizhnik-Zamolodchikov equation of level $0$. 
It is not clear at this point what kind of local operator 
corresponds to our solution. This is a general disadvantage 
of the axiomatic approach. 

In our integral formulae, the freedom in the solutions to 
the quantum Knizhnik-Zamolodchikov equation of level $0$ 
corresponds to the choice of the integral kernel 
$\Psi_{n}^{(\sigma)}(w|\zeta )$. In order to determine 
the structure of the space of form factors, 
we should study 
the annihilation pole condition for our integral formulae. 

It is very important to construct integral formulae for form 
factors of the XYZ antiferomagnet. If we choose the local operator 
${\cal O}$ as the identity operator, this can be done in a 
manner similar to that in Ref. \cite{qXYZ}. 
In other nontrivial cases this 
problem is not so easy. In connection with this problem, 
Lashkevich \cite{La} found 
a prescription to obtain these integral formulae 
in the vertex operator formalism. In principle, 
all form factors corresponding to all local fields 
can be constructed, but they take very complicated forms. 
We hope to express XYZ form factors in simpler forms 
on the basis of the axiomatic approach in a separate paper. 

\section*{Acknowledgements}
We would like to thank H. Konno and J. Shiraishi for 
their interest in this paper. 
This work is partly supported by a Grant-in-Aid for 
Scientific Research from JSPS (No. 15540218). 

%

\appendix

\section{Proof of Proposition \ref{prop:qKZ}} 

In this appendix, we prove (\ref{eq:W'-symm-comp}) and 
(\ref{eq:F'cyc-comp}) in order to prove Proposition \ref{prop:qKZ}. 

\subsection{Proof of $W'$-symmetry} 

Let us first prove (\ref{eq:W'-symm-comp}). For 
this purpose, we have to consider four cases, 
corresponding to $l_j -l_{j-1}=\pm 1$ and 
$l_{j+1}-l_j =\pm 1$. 

Suppose that $l_j -l_{j-1}=l_{j+1}-l_j =1$. Then the relation 
(\ref{eq:W'-symm-comp}) follows from (\ref{eq:g*-prop}), 
because $j,j+1\not\in A_-$, 
and consequently the integrand 
$\overline{F}^{(\sigma)}_n ({\cal O}; \zeta )$ is 
symmetric with respect to $\zeta_j$ and $\zeta_{j+1}$. 

When $(l_{j-1}, l_j , l_{j+1})=(m,m-1,m)$ for some $m$, 
the relation 
(\ref{eq:W'-symm-comp}) reduces to 
\begin{eqnarray}
&&\overline{F}_{n}^{(\sigma)} 
({\cal O}; \cdots,\zeta_{j+1},\zeta_j,\cdots)_{\cdots 
mm-1m \cdots} \nonumber \\[3mm]
&=&\dfrac{[1]'\{m-u_j +u_{j+1}\}'}{[1-u_j +u_{j+1}]'\{m\}'}
\overline{F}_{n}^{(\sigma)} 
({\cal O}; \cdots,\zeta_j,\zeta_{j+1},\cdots)_{\cdots 
mm-1m \cdots} \nonumber \\[5mm]
&+&\dfrac{[u_j -u_{j+1}]'\{m-1\}'}{[1-u_j -u_{j+1}]'\{m\}'}
\overline{F}_{n}^{(\sigma)} 
({\cal O}; \cdots,\zeta_j,\zeta_{j+1},\cdots)_{\cdots 
mm+1m \cdots}. 
\label{eq:W-symm/-+}
\end{eqnarray}
Note that the set of integration variables in 
the second term on the RHS is different from that of the 
other terms. Because the variables $w_a =x^{-2v_a}$ are 
the integration variables, 
we can replace both $v_{j}$ and $v_{j+1}$ in the integrand 
by $v$. Then, the relation (\ref{eq:W-symm/-+}) is obtained 
by equating the integrands. In this step, 
we use 
$$
\begin{array}{rcl}
\dfrac{\{v-u_{j+1}-\tfrac{1}{2}-m\}'}{
[u_{j+1}-v+\frac{3}{2}]'}&=&
\dfrac{[1]'\{m-u_j +u_{j+1}\}'}{[1-u_j +u_{j+1}]'\{m\}'}
\dfrac{\{v-u_{j}-\tfrac{1}{2}-m\}'}{
[u_j -v+\frac{3}{2}]'} \\[5mm]
&+&\dfrac{[u_j -u_{j+1}]'\{m-1\}'}{[1-u_j -u_{j+1}]'\{m\}'}
\dfrac{\{v-u_{j+1} -\tfrac{3}{2}-m\}'}{
[u_{j+1}-v+\tfrac{3}{2}]'}
\dfrac{[v-u_{j}-\tfrac{1}{2}]'}{
[u_{j}-v+\frac{3}{2}]'}. 
\end{array}
$$

Suppose that $(l_{j-1},l_j,l_{j+1})=(m,m+1,m)$ for some 
$m$. The proof in this case can be done 
in a way similar to that in the previous case. Here we use 
$$
\begin{array}{l}
\dfrac{\{v-u_{j}-\tfrac{3}{2}-m\}'}{
[u_{j}-v+\frac{3}{2}]'}\dfrac{[v-u_{j+1}-\tfrac{1}{2}]'}{
[u_{j+1}-v+\frac{3}{2}]'}=
\dfrac{[u_j -u_{j+1}]'\{m+1\}'}{[1-u_j -u_{j+1}]'\{m\}'}
\dfrac{\{v-u_{j}-\tfrac{1}{2}-m\}'}{
[u_j -v+\frac{3}{2}]'} \\[5mm]
\qquad\qquad +\dfrac{[1]'\{m+u_j -u_{j+1}\}'}{[1-u_j +u_{j+1}]'\{m\}'}
\dfrac{\{v-u_{j+1} -\tfrac{3}{2}-m\}'}{
[u_{j+1}-v+\tfrac{3}{2}]'}
\dfrac{[v-u_{j}-\tfrac{1}{2}]'}{
[u_{j}-v+\frac{3}{2}]'}. 
\end{array}
$$

Finally, let $(l_{j-1},l_j,l_{j+1})=(m,m-1,m-2)$ 
for some $m$. Then the integrand 
on the RHS of (\ref{eq:W'-symm-comp}) contains the factor 
$$
\begin{array}{cl}
& I(u_j,u_{j+1};v_j,v_{j+1}) \\
=&\dfrac{\{v_j-u_{j}-\frac{1}{2}-m\}'}{
[u_{j}-v_j+\frac{3}{2}]'}\dfrac{
\{v_{j+1}-u_{j+1}+\frac{1}{2}-m\}'}{
[v_{j+1}-u_{j+1}+\frac{3}{2}]'}\dfrac{
[v_{j+1}-u_j-\frac{1}{2}]'}{
[u_{j}-v_{j+1}+\frac{3}{2}]'}[v_j-v_{j+1}+1]'. 
\end{array}
$$
The corresponding factor on the LHS should be equal to 
$I(u_{j+1},u_j ;v_j,v_{j+1})$. 
Thus, the difference between the two sides 
contains the factor 
$$
\begin{array}{cl}
& I(u_j,u_{j+1};v_j,v_{j+1})-
I(u_{j+1},u_{j};v_j,v_{j+1})=\{m-1\}'\times \\
\times&[v_j-v_{j+1}-1]'[v_j-v_{j+1}+1]'
[u_{j}-u_{j+1}]'\{v_j +v_{j+1}-u_j -u_{j+1}-m-1\}', 
\end{array}
$$
which is symmetric with respect to $w_j =x^{-2v_j}$ and 
$w_{j+1}=x^{-2v_{j+1}}$. Because 
$\Psi^{(\sigma)}_n (w|\zeta )$ is antisymmetric 
with respect to the variables $w_a$, the relation 
(\ref{eq:W'-symm-comp}) in this case does hold. 

\subsection{Proof of the cyclicity} 

In the proof of the cyclicity (\ref{eq:F'cyc-comp}) 
we have to consider the two cases $l_{2n-1}=l\pm 1$. 
First, let $l_{2n-1}=l-1$. Note that 
$F_{n}^{(\sigma)}({\cal O}; \zeta)_{l\cdots l-1l}$ 
actually has no pole at the point $w_a =x^{-3}z_{2n}$, 
because $2n\not\in A_-$. 
When the integral (\ref{eq:G-form}) 
is analytically continued from 
$\zeta_{2n}=x^{-u_{2n}}$ to 
$x^{-2} \zeta_{2n}=x^{-(u_{2n}+2)}$, the points 
$x^{1+4n_1+2(r-1)n_2}z_{2n}$ and 
$x^{-1-4n_1-2(r-1)n_2}z_{2n}$ 
($n_1 , n_2 \in\mathbb{Z}_{\geqslant 0}$) 
move to the points 
$x^{-3+4n_1+2(r-1)n_2}z_{2n}$ and 
$x^{-5-4n_1-2(r-1)n_2}z_{2n}$, respectively. 
Thus, the integral contour 
$C'_a=C_a (z', x^{-4} z_{2n})$ coincides with 
the original one: $C_a =C_a (z)$. 

Furthermore, using (\ref{eq:trPsi1}) we obtain 
$$
\Psi^{(\sigma)}_n (w|\zeta', x^{-2} \zeta_{2n})
=\sigma\Psi^{(\sigma)}_n (w|\zeta)
\prod_{j=1}^{2n-1} \left(\frac{\zeta_j}{\zeta_{2n}} 
\right)^{\frac{r}{r-1}}
\prod_{a\in A_-} \dfrac{[v_a -u_{2n}-\tfrac{1}{2}]'}{
[u_{2n} -v_a +\tfrac{3}{2}]'}, 
$$
which implies that the integrands of the two sides of 
(\ref{eq:F'cyc-comp}) are identified, and therefore the 
relation (\ref{eq:F'cyc-comp}) holds when $l_{2n-1}=l-1$. 

Next, let $l_{2n-1}=l+1$. 
In this case we rescale the 
variable $w_{2n}$ as $w_{2n}\mapsto x^{-4}w_{2n}$ 
($v_{2n}\mapsto v_{2n}+2$) on the LHS of 
(\ref{eq:F'cyc-comp}). Then, the integral 
contour with respect to $w_a$ ($a\in A_-\backslash \{ 2n\}$) 
will be $C'_a =C_a (z', x^{-4} z_{2n})=C_a (z)$, 
for the same reason as in the previous case. 
The other contour, 
$\widetilde{C}=C_{2n}(x^{4}z', z_{2n})$, 
encircles $x^{5+4n_1+2(r-1)n_2}z_j$, 
$x^{1+2(r-1)n_3}z_{j}$ 
and $x^{1+4n_1+2(r-1)n_2}z_{2n}$, 
$x^{-3+2(r-1)n_3}z_{2n}$, 
but not 
$x^{3-4n_1-2(r-1)n_2}z_j$, $x^{1-2(r-1)(n_3+1)}z_{j}$ 
nor $x^{-1-4n_1-2(r-1)n_2}z_{2n}$, 
$x^{-3-2(r-1)(n_3+1)}z_{2n}$, 
where $1\leqslant j\leqslant 2n-1$ and 
$n_1 , n_2 , n_3\in \mathbb{Z}_{\geqslant 0}$. 

Since 
$Q^{(n)}(w',x^{-4}w_{2n}|\zeta' , 
x^{-2} \zeta_{2n})_{l\cdots l+1l}$ 
contains the factor 
\begin{equation}
\begin{array}{cl}
& J(w', x^4 w_{2n}|\zeta',x^2\zeta_{2n}) \\
=&\displaystyle\dfrac{
\{v_{2n}-u_{2n}-\tfrac{3}{2}-l\}'}{
[u_{2n}-v_{2n}+\tfrac{3}{2}]'}
\prod_{j=1}^{2n-1} \dfrac{[v_{2n}-u_j+\tfrac{3}{2}]'}{
[u_{j}-v_{2n} -\tfrac{1}{2}]'}
\prod_{a\in A_-\atop a\neq 2n} 
[v_a -v_{2n}-1]', \end{array}
\label{eq:Qtr}
\end{equation}
the pole at $w_{2n}=x^{3}z_{j}$ ($1\leqslant j\leqslant 2n-1$) 
disappears. Thus, we can deform the contour 
$\widetilde{C}$ so as to coincide with 
the original one, $C_{2n}=C_{2n}(z)$, 
without crossing any poles. Thus, the integral contours 
on the two sides of (\ref{eq:F'cyc-comp}) coincide. 

Next, replace the integral variables as 
$(w', w_{2n})\mapsto (w_{2n}, w')$ on the RHS 
of (\ref{eq:F'cyc-comp}) and compare the integrands of the two 
sides. Then, the corresponding factor on the RHS of (\ref{eq:Qtr}) 
is equal to 
\begin{equation}
J(w_{2n}, w'|\zeta ) =
\displaystyle\dfrac{
\{v_{2n}-u_{2n}-\tfrac{3}{2}-l\}'}{
[u_{2n}-v_{2n}+\tfrac{3}{2}]'}
\prod_{a\in A_-\atop a\neq 2n} 
\dfrac{[v_{a}-u_{2n}-\tfrac{1}{2}]'}{
[u_{2n}-v_{a} +\tfrac{3}{2}]'} 
[v_{2n}-v_a +1]'. 
\label{eq:Qcyc}
\end{equation}
{}From (\ref{eq:Qtr}), (\ref{eq:Qcyc}), 
(\ref{eq:trPsi1}), (\ref{eq:trPsi2}) 
and the antisymmetric property of $\Psi^{(\sigma)}_n$ 
with respect to the variables $w_a$, we have 
\begin{equation}
\begin{array}{cl}
&\Psi^{(n)}_\sigma (w', x^{-4} w_{2n}|
\zeta', x^{-2} \zeta_{2n})
J(w', x^{-4} w_{2n}|\zeta',x^{-2}\zeta_{2n}) \\
=&\displaystyle\sigma \Psi^{(\sigma)}_n 
(w_{2n}, w'|\zeta ) J(w_{2n}, w'|\zeta )
\prod_{j=1}^{2n-1} \left( \frac{\zeta_j}{\zeta_{2n}} 
\right)^{\frac{r}{r-1}}, \end{array}
\label{eq:Psi12}
\end{equation}
which implies that the integrands of the two sides of 
(\ref{eq:F'cyc-comp}) are identical, and therefore the relation 
(\ref{eq:F'cyc-comp}) holds when $l_{2n-1}=l+1$.


\begin{thebibliography}{99}
  
\bibitem{qXYZ}Y.-H. Quano, Bootstrap equations and 
correlation functions for the Heisenberg XYZ antiferromagnet, 
J. Phys. A: Math Gen {\bf 35} (2002) 9549--9572. 
\bibitem{JMN}M. Jimbo, T. Miwa and A. Nakayashiki, 
Difference equations for the correlation functions of
the eight-vertex model, {\it J. Phys. A: Math. Gen.} 
{\bf 26} (1993) 2199--2209. 
\bibitem{Smbk}F. A. Smirnov, {\it 
Form factors in completely integrable models of
quantum field theory}, Advanced Series in Mathematical 
Physics Vol {\bf 14}, (World Scientific, Singapore, 1992).
\bibitem{Sm1}F. A. Smirnov, Dynamical 
symmetries of massive integrable models 1,2, 
in {\it Proceedings of the RIMS Research Project 1991, 
``Infinite Analysis''}; 
{\it Int. J. Mod. Phys.} {\bf 7A, Suppl. 1B} (1992) 
813--837; 839--858. 
\bibitem{FR}I. B. Frenkel and N. Y. Reshetikhin,
Quantum affine algebras and holonomic difference equations,
{\it Commun. Math. Phys.} {\bf 146} (1992) 1--60. 
\bibitem{JMbk}M. Jimbo and T. Miwa, 
{\it Algebraic analysis of solvable lattice models}, 
CBMS Regional Conferences Series in Mathematics 
Vol {\bf 85}, (AMS, Providence, RI, 1994). 
\bibitem{Lu}S. Lukyanov, Free field representation 
for massive integrable models, 
{\it Commun. Math. Phys.} {\bf 167} (1995)183--226. 
\bibitem{La}M. Lashkevich, Free field construction 
for the eight-vertex model: representation for form factors, 
{\it Nucl. Phys.} {\bf B621} (2002) 587-621. 
\bibitem{3-bu/2}R. J. Baxter, Eight-vertex model in lattice 
statistics and one-dimensional anisotropic 
Heisenberg chain. II. 
Equivalence to a generalized ice-type lattice model, 
{\it Ann. Phys. (NY)} {\bf 76} (1973) 25--47. 
\bibitem{LuP}S. Lukyanov and Ya Pugai, 
Multi-point local height probabilities in the 
integrable RSOS model, 
{\it Nucl. Phys.} {\bf B473}[FS] (1996) 631--658. 
\bibitem{MW}T. Miwa and R. Weston, 
Boundary ABF Models, {\it Nucl. Phys.} {\bf B486 [PM]} 
(1997) 517--545. 
\bibitem{LaP}M. Lashkevich and Ya Pugai, 
Free field construction for correlation 
functions of the eight vertex model, 
{\it Nucl. Phys.} {\bf B516} (1998) 623--651. 
\bibitem{Shi}J. Shiraishi, Free field constructions for 
the elliptic algebra ${\cal A}_{q,p}(\widehat{\mathfrak{sl}_2})$ 
and Baxter's eight-vertex model, math.QA/0302097, 2003. 
\bibitem{SKAO}J. Shiraishi, H. Kubo, H. Awata and 
S. Odake, A quantum deformation of the Virasoro algebra 
and Macdonald symmetric functions, 
{\it Lett. Math. Phys.} {\bf 38} (1996) 33--57. 
\bibitem{S1}F. A. Smirnov, Counting the local fields in sine-Gordon theory, 
Nucl. Phys. {\bf B453} [FS] 807--824. 
\bibitem{BBS2}O. Babelon, D. Bernard, and F. A. Smirnov, 
Null-vectors in integrable field theory, 
{\it Commun. Math. Phys.} {\bf 186} (1997) 601--648. 
\bibitem{JKMQ}M. Jimbo, T. Kojima, T. Miwa 
and Y.-H. Quano, Smirnov's integral and quantum 
Knizhnik-Zamolodchikov equation of level $0$, 
{\it J. Phys. A: Math. Gen.} {\bf 27} (1994) 3267--3283. 
\bibitem{KMQ}T. Kojima, K. Miki and Y.-H. Quano, 
Annihilation poles of a Smirnov-type integral formula 
for solutions to the quantum Knizhnik--Zamolodchikov
equation, J. Phys. A: Math. Gen. {\bf 28} 3479--3491. 
\bibitem{affine}Y.-H. Quano, Form factors of the 
XXZ model and quantum affine symmetry, 
{\it J. Phys. A: Math. Gen.} {\bf 31} (1998) 
1791--1800.  
\bibitem{NT}A. Nakayashiki and Y. Takeyama, 
On form factors of SU(2) invariant Thirring model, 
{\it MathPhys Odyssey 2001, Integrable Systems and Beyond}, 
in Honor of Barry M. McCoy, M. Kashiwara and T. Miwa eds., 
Progress in Mathematical Physics {\bf 23} 357--390, (2002) 
Birkh\"{a}user. 
\bibitem{N3}A. Nakayashiki, 
Residues of $q$-hypergeometric integrals and characters 
of affine Lie algebras, {\it Commun. Math. Phys.} (2003) {\bf 197} 
197--241. 
\bibitem{Matsuo}A. Matsuo, Quantum algebra 
structure of certain Jackson integrals, 
Comm. Math. Phys. {\bf 157} (1993) 479-498. 
\bibitem{TV}V. Tarasov and A. Varchenko, 
Jackson integral representations for solutions of the quantized
Knizhnik--Zamolodchikov equation, 
{\it St. Petersburg Math. J}. {\bf 6} (1995) 275--313. 
\bibitem{NPT}A. Nakayashiki, A. Pakuliak and V. Tarasov, 
On solutions of the KZ and qKZ equations at level zero, 
{\it Ann. Inst. Henri Poincar\'{e}}, {\bf 71} (1999) 459--496. 
\bibitem{KMT0}N. Kitanine, J. M. Maillet and V. Terras, 
Correlation functions of the XXZ Heisenberg spin-1/2
chain in a magnetic field, {\it Nucl. Phys.} {\bf B567} 
(2000) 554-582. 
\bibitem{KMST1}N. Kitanine, J. M. Maillet, N. A. Slavnov 
and V. Terras, Spin-spin correlation functions of the XXZ-1/2 Heisenberg 
chain in a magnetic field, {\it Nucl. Phys.} 
{\bf B641} (2002) 487-518. 
\bibitem{dYBE}G. Felder, Elliptic quantum groups, 
in {\it Proc. ICMP Paris} pp211--218, 1994. 
\end{thebibliography}
\end{document}